\documentclass[9pt,twocolumn,twoside]{optica}
\setboolean{shortarticle}{false}
\setboolean{minireview}{false}
\usepackage{hyperref}

\title{Scattering of Poincar\'e beams: polarization speckles}

\author[1, *]{Salla Gangi Reddy}
\author[2]{Vijay Kumar}
\author[1]{Yoko Miyamoto}
\author[2]{R. P. Singh}

\affil[1]{The University of Electro-Communications, Chofugaoka, Chofu, Tokyo - 1828585, Japan.}
\affil[2]{Physical Research Laboratory, Navarangpura, Ahmedabad-380009, India.}

\affil[*]{Corresponding author: sgreddy@uec.ac.jp}

\dates{Compiled \today}

\ociscodes{(260.5430) Polarization; (030.1640) Coherence; (030.6140) Speckle; (050.4865) Optical Vortices.}

\doi{\url{http://dx.doi.org/10.1364/optica.XX.XXXXXX}}

\begin{abstract}
Polarization speckle is a fine granular light pattern having spatially varying random polarization profile. We generate these speckle patterns by using the scattering of Poincar\'e beams, a special class of vector vortex beams, through a ground glass plate. Here, the Poincar\'e beams are generated using a polarization sensitive spatial light modulator displaying an on-axis hologram corresponding to an optical vortex phase profile. The different inhomogeneities of the rough surface experience different polarizations, which control the ability for scattered waves to interfere at the detection plane and causes a spatially varying polarization profile. We experimentally determined the spatial variation of local degree of polarization and orientation of the polarization ellipse for these speckle patterns from the Stokes analysis. We also determined the size of scalar speckles using the auto-correlation function of Stokes parameter $S_0$ and the size of polarization speckles using the sum of auto-correlation functions of remaining three Stokes parameters. We found that the scalar speckle size is independent of index of the vector beam but the size of polarization speckles decreases with the increase in index of the vector beam.
\end{abstract}

\setboolean{displaycopyright}{true}

\begin{document}

\maketitle
\thispagestyle{fancy}
\ifthenelse{\boolean{shortarticle}}{\abscontent}{}

\section{Introduction}

The cylindrical vector beams are known for their non-separability in polarization and spatial mode as well as for spatially varying polarization states \cite{vector1, vector2}. These beams are solutions to the vector wave equation and have a vectorial singularity in their polarization distribution as a C-point at which the orientation of polarization ellipse  is undefined or L-line at which the sign of rotation is undefined \cite{vector3, vector4, vector5}. If the vectorial singularity coincides with optical phase singularity, then the beams are called vector vortex beams. These beams have ability to produce tighter focal spots and stronger longitudinal field gradients \cite{vector6}. The non-separability of vector beams is utilized in many applications such as polarization metrology \cite{vector7}, high speed kinematic sensing \cite{vector8} and single shot Mueller polarimetry \cite{vector9}. The vector beams can be generated using laser resonators\cite{vector10}, 
liquid crystal display elements \cite{vector11} and also using the superposition of Laguerre-Gaussian beams \cite{vector12}. The vector vortex beams generated by the superposition of LG beams contain the polarization states that completely span the surface of Poincar\'e sphere and named Poincar\'e beams. The polarization singularities are also realized in random light fields such as speckles \cite{vector13, vector14, vector15}. 
  
Speckle pattern is a fine granular pattern of light obtained by the scattering of a coherent light beam \cite{vector16}. These patterns are due to the interference of many scattered wave fronts coming from different scatterers of a rough surface to the detection plane. The speckles and their statistical properties are utilized in a variety of applications such as imaging, optical coherence tomography, and astronomy \cite{vector17}. We have previously studied the scattering of optical vortices in detail and found that the temporal intensity correlation for scattered light decays sharper with the order of a vortex \cite{vector18}. It is also observed that the size of speckles decreases with the order that is attributed to the increase in area of illumination at the scattering plane due to increase in order of the vortex \cite{vector19, vector20}. Recently, the non-diffracting nature has been realized for the random fields generated by scattering the "perfect" optical vortices, whose size and shape are invariant with the order \cite{vector21}. The speckles with spatially varying polarization i.e. polarization speckles, are getting a lot of attention in biomedical optics especially for distinguishing different types of skin cancers \cite{vector22, vector25}. These speckles can be generated using the superposition of two speckle patterns obtained by the scattering of orthogonally polarized Gaussian beams through  statistically independent random media \cite{vector23}. 

In this work, we demonstrate a scheme for generating polarization speckles using the scattering of Poincar\'e beams of different indices. When we scatter these beams each and every scatterer experiences different polarization and all of the scattered waves with random polarizations interfere at the detection plane resulting in the formation of polarization speckles. We also present the statistical properties of generated speckles using Stokes analysis and quantification of scalar and vector speckle size. It has been showed that the scalar speckle size is independent of the index of Poincar\'e beams and the size of polarization speckles decreases with the increase in index.

\section{Experimental details}
\label{sec:expt}
The experimental set up for generating the vector beams as well as polarization speckles is shown in Fig. (1). A diode pumped solid state laser (Coherent Verdi-V10) of wavelength 532 nm with vertical polarization has been used. The laser beam passes through a half wave plate (HWP) in order to change the composition of two orthogonal polarizations i.e. horizontal (H) and vertical (V). The spatial light modulator (SLM) used here is sensitive to polarization i.e. the displayed phase will be imparted to the vertically polarized light only and the light with horizontal polarization will be unaffected.
\begin{figure}[h]
\begin{center}
\includegraphics[width=3.0in]{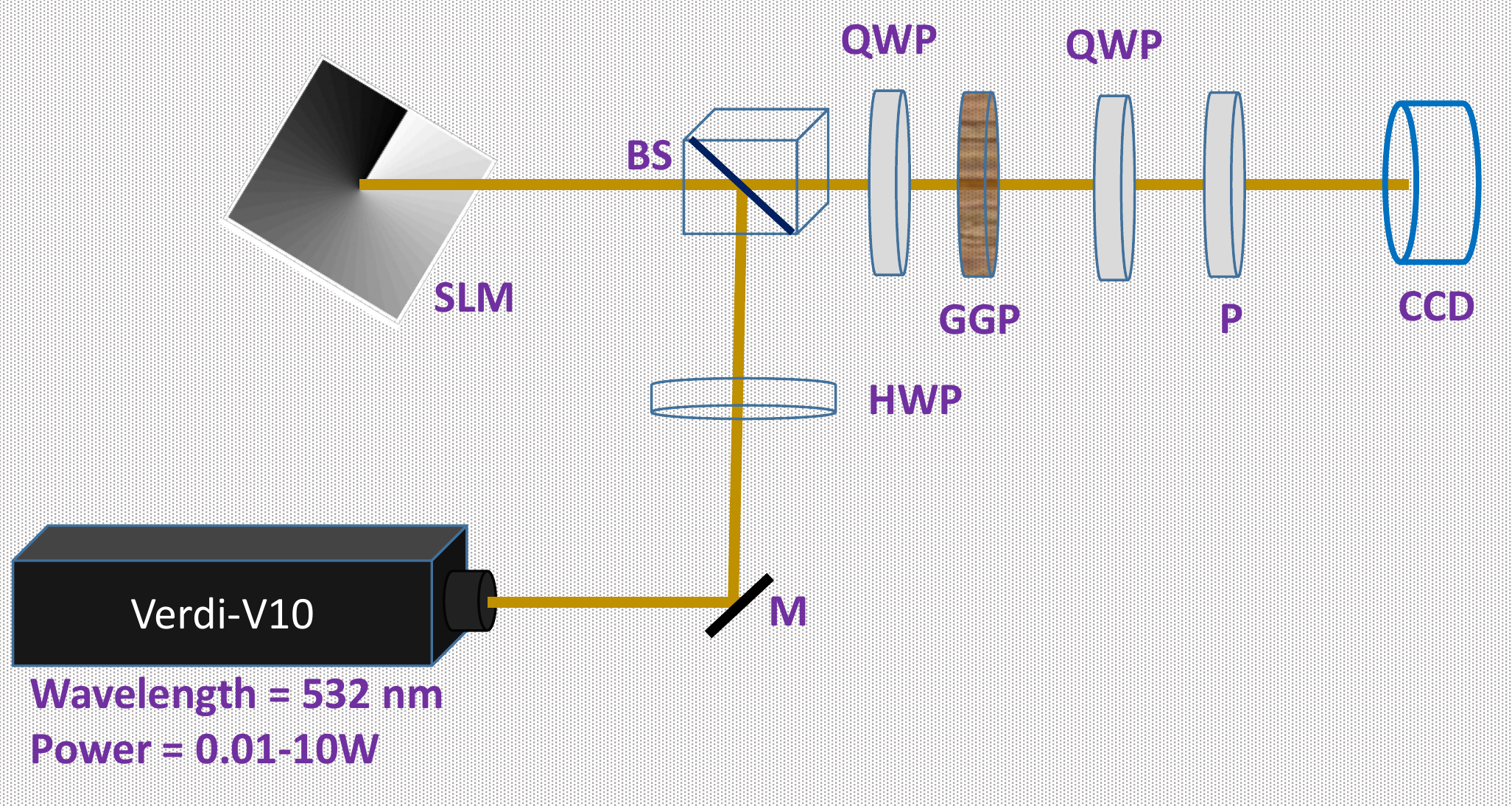}
\caption{ (Colour online) The experimental set-up for the generation of Poincar\'e beams and their scattering to generate polarization speckles. Here M--Mirror, HWP--Half wave plate, BS--Beam splitter, SLM--Spatial light modulator, QWP--Quarter wave plate, GGP--Ground glass plate, P--Polarizer, CCD--Camera.}\label{fig:expt}
\end{center}
\end{figure}

 \begin{table*}   
 \begin{center}
  \begin{tabular}{|c|c|c|c|c|} \hline
\hspace{0.3cm} $\theta$ \hspace{0.3cm} & \hspace{0.7cm} Polarization on SLM \hspace{0.7cm}  & \hspace{0.7cm} $c_1$   \hspace{0.7cm}    &  \hspace{0.7cm} $c_2$ \hspace{0.7cm}  &  \hspace{0.7cm} Nature of the beam \\ \hline
\hspace{0.3cm} $0^\circ$ \hspace{0.3cm} & \hspace{0.7cm} Vertical  \hspace{0.7cm}  & \hspace{0.7cm} 0   \hspace{0.7cm}    &  \hspace{0.7cm} 1 \hspace{0.7cm}  &  \hspace{0.7cm} Scalar vortex beam \\ \hline
\hspace{0.3cm} $22.5^\circ$ \hspace{0.3cm} & \hspace{0.7cm} Diagonal  \hspace{0.7cm}  & \hspace{0.7cm} $\frac{1}{\sqrt{2}}$   \hspace{0.7cm}    &  \hspace{0.7cm}$\frac{1}{\sqrt{2}}$ \hspace{0.7cm}  &  \hspace{0.7cm} Vector vortex beam \\ \hline
\hspace{0.3cm} $45^\circ$ \hspace{0.3cm} & \hspace{0.7cm} Horizontal  \hspace{0.7cm}  & \hspace{0.7cm} 1   \hspace{0.7cm}    &  \hspace{0.7cm} 0 \hspace{0.7cm}  &  \hspace{0.7cm} Scalar Gaussian beam \\ \hline
\end{tabular}                                
\caption{ The nature of the generated beam and the incident polarization states for different fast axis orientations of HWP.} 
\end{center}
\end{table*}

Here, we display an on-axis hologram for generating the optical vortex beam. For on-axis holograms, the phase pattern of vortex beams is displayed on the SLM and the singularity in phase is directly transferred to the vertically polarized Gaussian beam so that the vertically polarized vortices are generated in the central order and overlap with the unaffected H component. The output from the SLM is the superposition of horizontally polarized Gaussian beam and vertically polarized vortex beam. This is the vector vortex beam in the linear polarization basis that contains two opposite C-singular points in its polarization profile for the case of topological charge 1 for the vortex \cite{vector4}. This beam can be converted to the vector vortex beam in circular polarization basis using a quarter wave plate (QWP) with an isolated C-singular point. This superposition forms a Poincar\'e beam containing either lemon or star patterns in the polarization ellipse orientation field. We also generate vector vortex beams of higher indices by displaying on-axis phase holograms corresponding to the higher order optical vortices on SLM. Poincar\'e beams are known for their non-separability in polarization and spatial mode. It can be represented mathematically as: 
\begin{equation}
|\Psi> = c_1|R,0>+c_2|L,m>
\end{equation}
where R, L represent the right circular and left circular polarizations. 0, \textit{m} represent the azimuthal mode indices of the spatial modes and $c_1$ and $c_2$ are superposition constants. The constants $c_1$ and $c_2$ can be easily controlled just by rotating the fast axis of HWP. The constants in terms of fast axis orientation of HWP are given by
 \begin{equation}
c_1 = \sin2\theta; c_2 = \cos2\theta
\end{equation}
The polarization of light incident on SLM and nature of the generated beam for different fast axis orientations of HWP have been summarized in table(1).

\section{Results and discussion}

\subsection{Polarization profile of vector beams}

\begin{figure}[h]
\begin{center}
\includegraphics[width=3.0in]{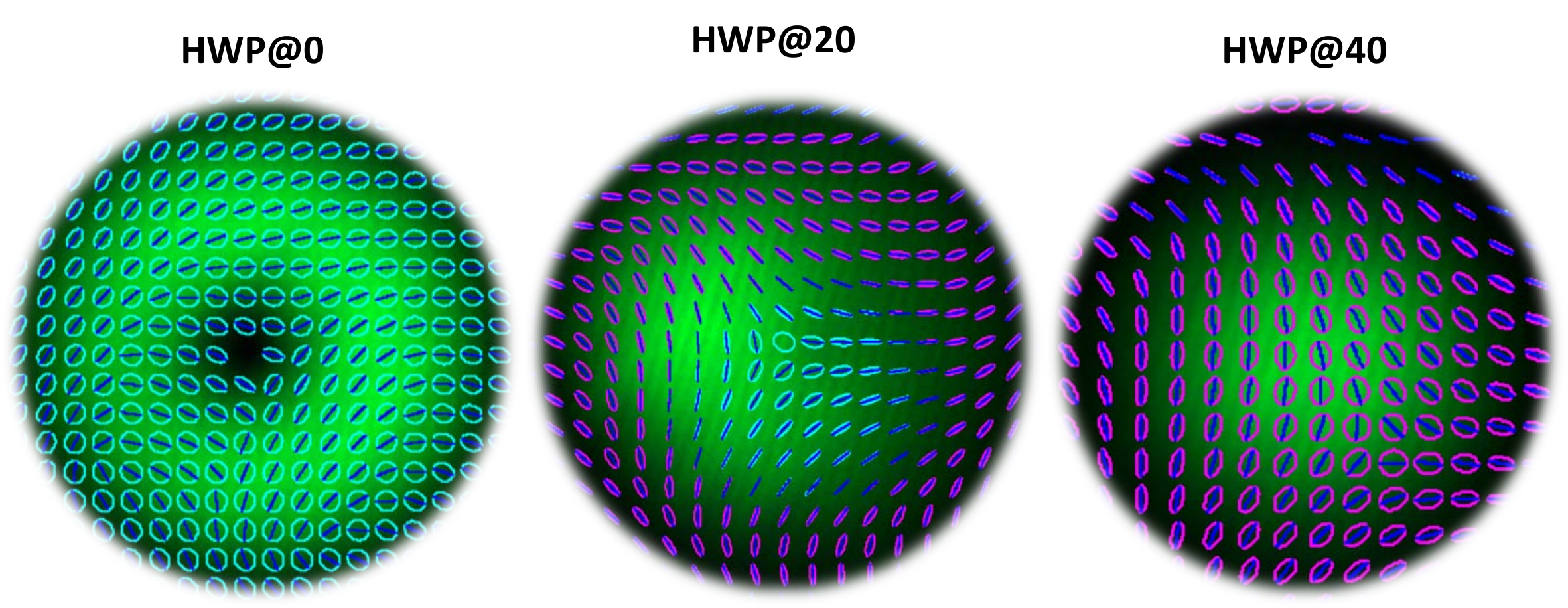}
\caption{ (Colour online) The experimentally generated polarization profiles of vector vortex beams for different HWP fast axis orientations $\theta$ for $m$=1. Green and magenta colour ellipses represent the right circular and left circular nature of the polarization states respectively.}\label{fig:fig2}
\end{center}
\end{figure}

Figure \ref{fig:fig2} shows the polarization profiles of generated vector beams with isolated C-singular point corresponding to the different fast axis orientations of HWP. We have studied the evolution of polarization singularity with the change in fast axis orientation of HWP that controls the amount of non-separability of the beam. At zero fast axis orientation of HWP, we have only vertical polarization incidents on SLM that forms the left circularly polarized scalar vortex beam. The same is verified (left in Fig.(\ref{fig:fig2})) using polarization profiles. At 45$^\circ$ fast axis orientation, we have only horizontal polarization which will be remain unaffected by SLM i.e. at output we have right circularly polarized Gaussian beam. We didn't observe the singularity in polarization at 40$^\circ$ fast axis orientation of the half wave plate, which may be due to the uncertainties in the angles of wave plates used to manipulate the polarization (right in Fig.(\ref{fig:fig2})).
However, at 22.5$^\circ$ fast axis orientation, the HWP converts vertical polarization into diagonal that has equal components for horizontal and vertical polarizations. After the SLM, we have equal amount of horizontally polarized Gaussian beam and vertically polarized vortex beam whose superposition forms the vector vortex beam in linear polarization basis. After passing through QWP, this beam converted as the vector vortex beam in circular polarization basis. We have shown (middle in Fig.(\ref{fig:fig2})) the formation of polarization singularity at 20$^\circ$ fast axis orientation of HWP. These polarization singularities are having star pattern which is consistent with the  $m\sigma$ sum rule \cite{vector3}, where $\sigma=\pm1$ is the spin angular momentum component.
\subsection{Polarization speckles}

\begin{figure}[h]
\begin{center}
\includegraphics[width=3.0in]{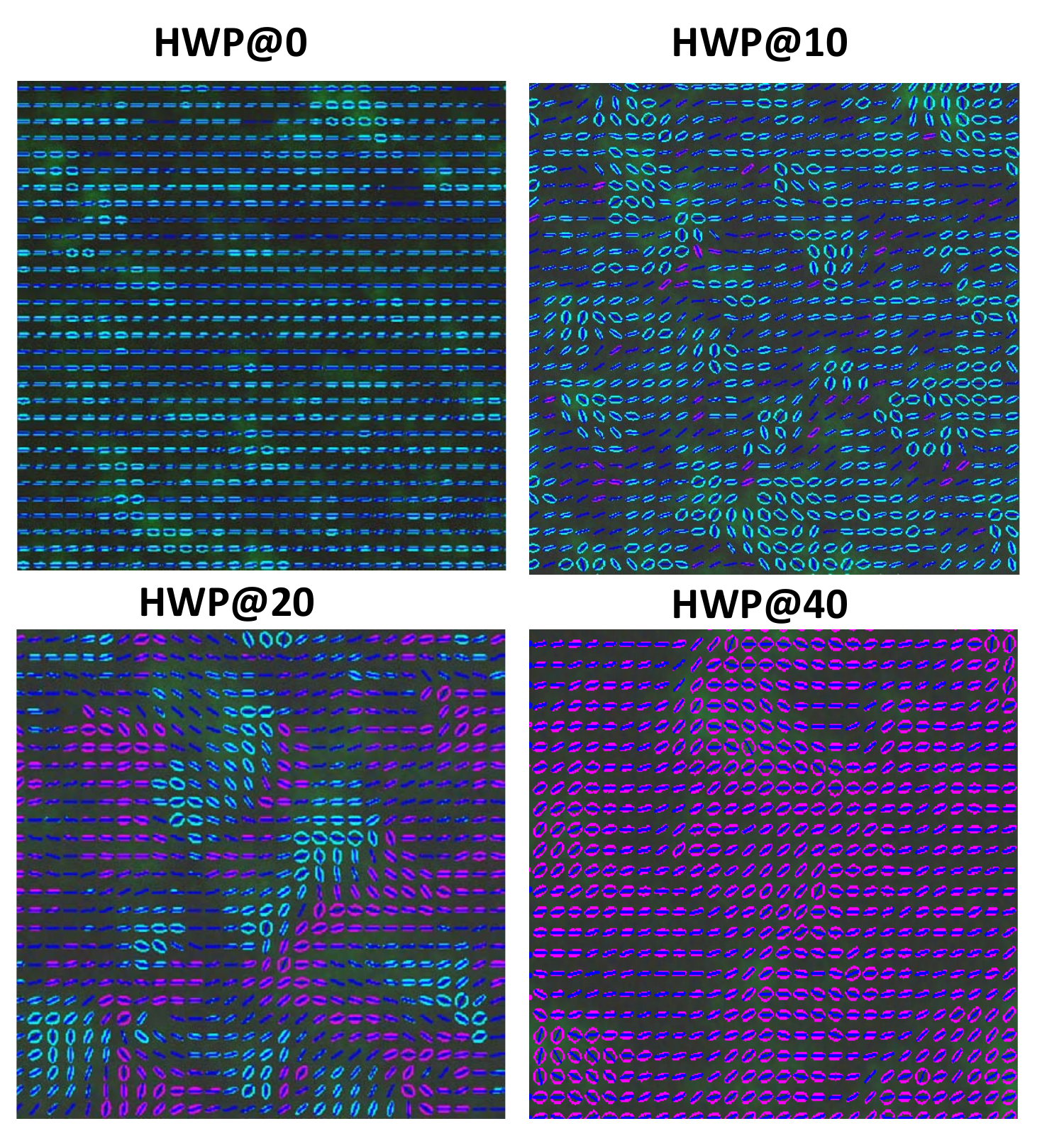}
\caption{ (Colour online) The experimental results for polarization speckles obtained by the scattering of vector vortex beam of order $m$=1 for different HWP fast axis orientations. Green and magenta colour ellipses represent the left circular and right circular nature of the polarization states respectively.}\label{fig:fig3}
\end{center}
\end{figure}

After verifying the presence of a star pattern in the polarization profile of vector vortex beam using the polarimetry, we scatter these beams through a rough surface, the ground glass plate (GGP) obtained from Thorlab (DG-20-220). The generated speckles have been recorded for different polarization projections using a QWP and a Polarizer in order to get the Stokes parameters. The spatial polarization structure of the speckles generated by the scattering of vector vortex beams at different fast axis orientations of the HWP are shown in Fig. \ref{fig:fig3}. The ellipses have been constructed using the Stokes parameters. For the zero degree fast axis orientation of HWP, the speckles have spatially uniform polarization as we are scattering the scalar vortex beam i.e. spatially uniform polarized light beam. However, with the increase in fast axis orientation of HWP, the superposition of Gaussian and vortex beams of different intensities will be formed. When the intensities of Gaussian and vortex beams are equal then the beam will have spatially varying polarization pattern to the maximum extent(at 22.5$^\circ$ fast axis orientation of HWP). When we scatter these non-separable beams, the generated speckles have spatially non-uniform i.e. random polarization profile as shown in Fig. \ref{fig:fig3}. The non-separability of polarization and spatial mode increases when we change the HWP orientation from 0$^\circ$ to 22.5$^\circ$. Due to the spatial variation of polarization in Poincar\'e beams, the different inhomogeneities experience different polarizations while scattering through the ground glass plate. The scattered waves having different polarizations interfere at the detection plane and form the resultant polarization profile which is random in nature and this randomness increases with the increase in the non-separability of the beam. 

\subsection{Stokes parameter distributions}

\begin{figure}[h]
\begin{center}
\includegraphics[width=3.0in]{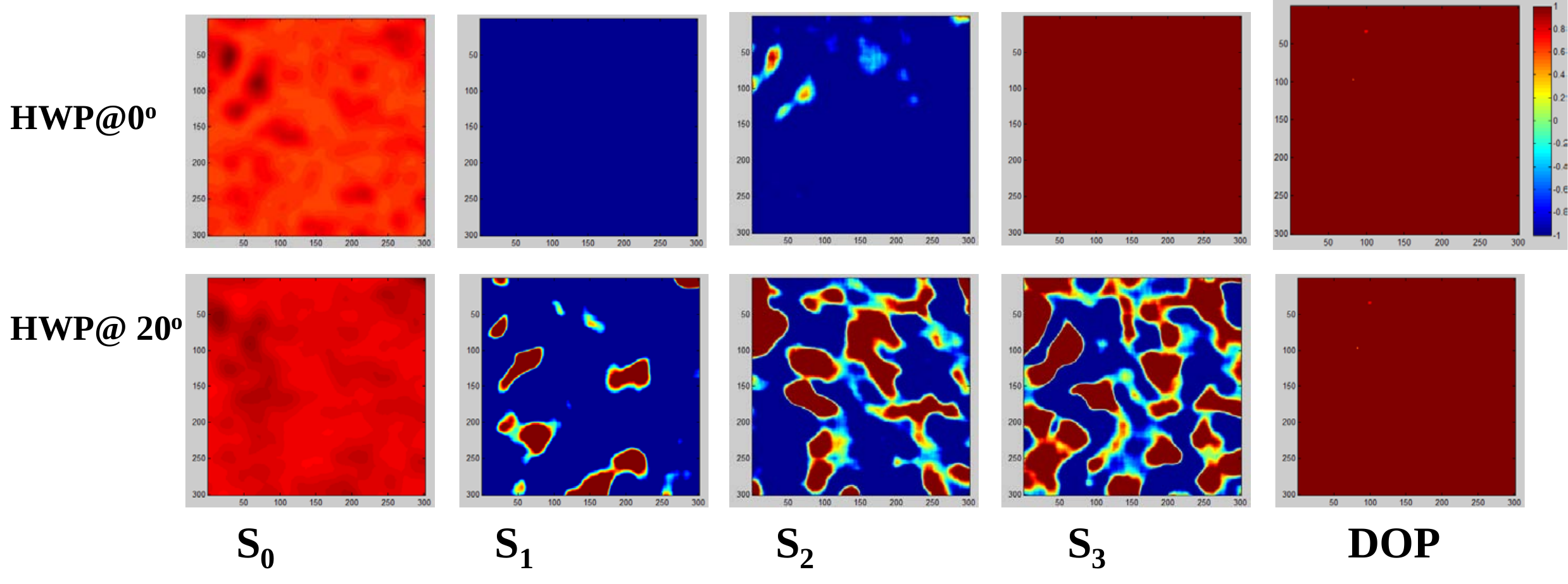}
\caption{ (Colour online) The spatial variation of Stokes parameters and degree of polarization (DOP) of polarization speckles corresponding to the order $m$=1 at HWP fast axis orientation of 0$^\circ$ (top) and 20$^\circ$ (bottom).}\label{fig:fig4}
\end{center}
\end{figure}

The spatial variation of polarization in the speckle pattern is confirmed through the Stokes parameter distributions. At 0$^\circ$ fast axis orientation of HWP, we have only circular polarization component, which is in agreement with polarization of the incident beam as it has uniform left circular polarization and shown at the top of Fig. \ref{fig:fig4}. At 20$^\circ$ fast axis orientation of HWP, the incident beam has spatially varying polarization and different inhomogeneities experience different polarization and produces random spatial polarization profile, consequently all Stokes parameters are random in space as shown in bottom of  Fig. \ref{fig:fig4}.

\begin{figure}[h]
\begin{center}
\includegraphics[width=3.2in]{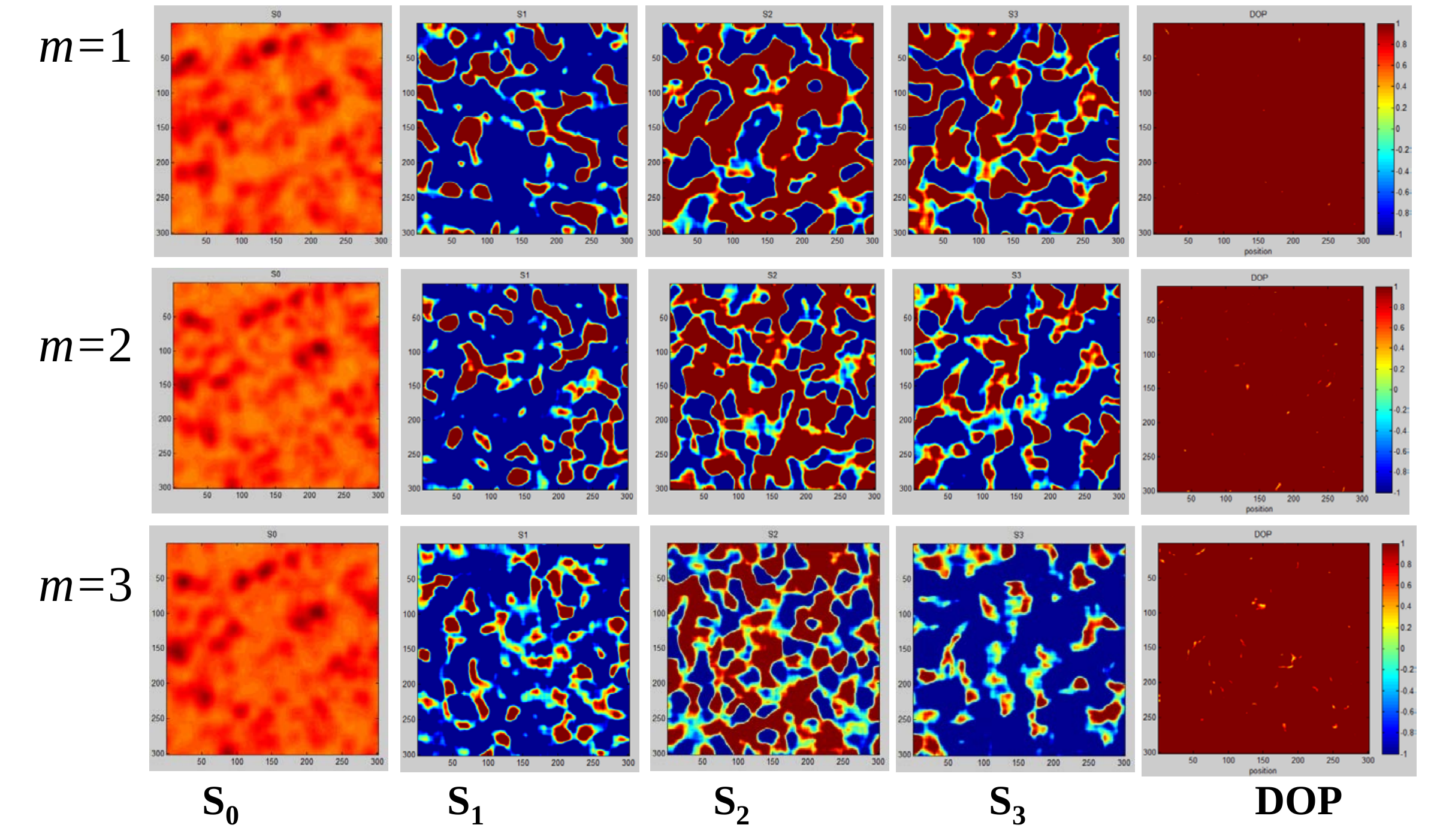}
\caption{ (Colour online) The Stokes parameters and DOP of polarization speckles corresponding to different indices of Poincar\'e beams at 20$^\circ$ fast axis orientation of the HWP.}\label{fig:fig5}
\end{center}
\end{figure}

We also verify the presence of random polarization for different indices of Poincar\'e beams which are shown in Fig. \ref{fig:fig5} while keeping the fast axis orientation of HWP as 20$^\circ$. It is clear from both the figures (\ref{fig:fig4}, \ref{fig:fig5})that the speckles have well defined polarization at each and every local point and degree of polarization is equal to 1. However, the spatial averaging over the entire space gives the degree of polarization as '0' corresponding to un-polarized light. One can also observe that the width of random polarization structures present in Stokes parameters decreases with the increase in index of Poincar\'e beams. This suggests that the width of auto-correlation function of Stokes parameters decreases with the index.   

\subsection{Size of polarization speckles}

We have also done a quantitative analysis in order to determine the size of scalar and vector polarization speckles, the latter being length scale up to which the correlations in polarization exist. This is done using the statistical properties of Stokes parameters. It has been suggested that the width of the auto-correlation function of Stokes parameter $S_0$ gives the scalar speckle size and the width of the sum of auto-correlation functions of three Stokes parameters can be considered the size of polarization speckles \cite{vector23}:
\begin{eqnarray}
C_s = <S_0S_0> \\
C_p = <S_1S_1>+<S_2S_2>+<S_3S_3>
\end{eqnarray}
where $C_p, C_s$ are the correlation function of polarization and scalar speckles, and $(S_0,S_1,S_2,S_3)$ are Stokes parameters. 

\begin{figure}[h]
\begin{center}
\includegraphics[width=3.0in]{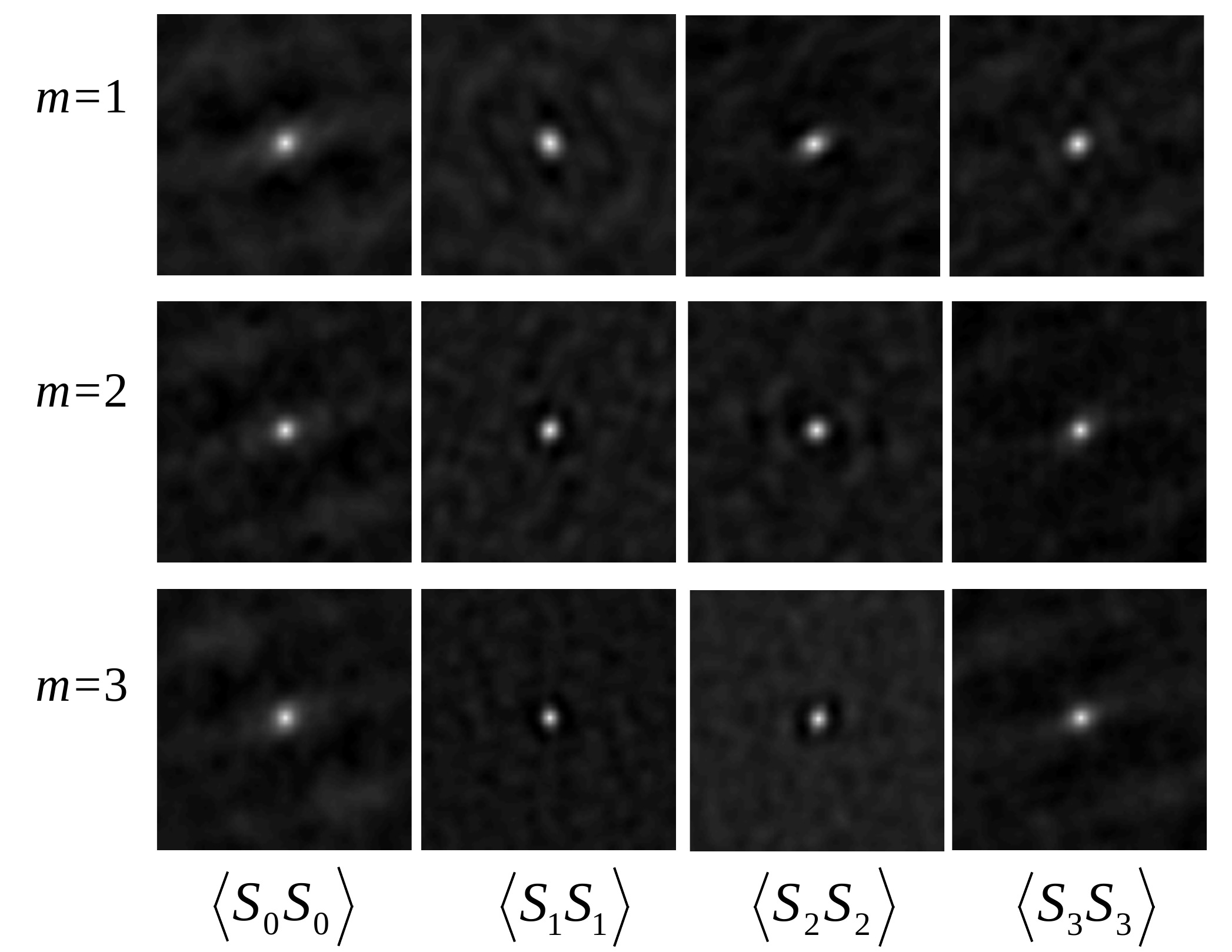}
\caption{The 2-D auto-correlation functions for the Stokes parameters.}\label{fig:fig6}
\end{center}
\end{figure}

Figure \ref{fig:fig6} shows the auto-correlation functions for all the four Stokes parameters corresponding to different indices of vector beam $m$ = 1, 2, 3 and at 20$^\circ$ fast axis orientation of HWP.
It seems from the figure that the size of scalar speckles is independent of the index whereas the size of polarization speckles decreases with the index of the vector beam. This has been confirmed by quantifying the width of correlation function. 

\begin{figure}[h]
\begin{center}
\includegraphics[width=3.0in]{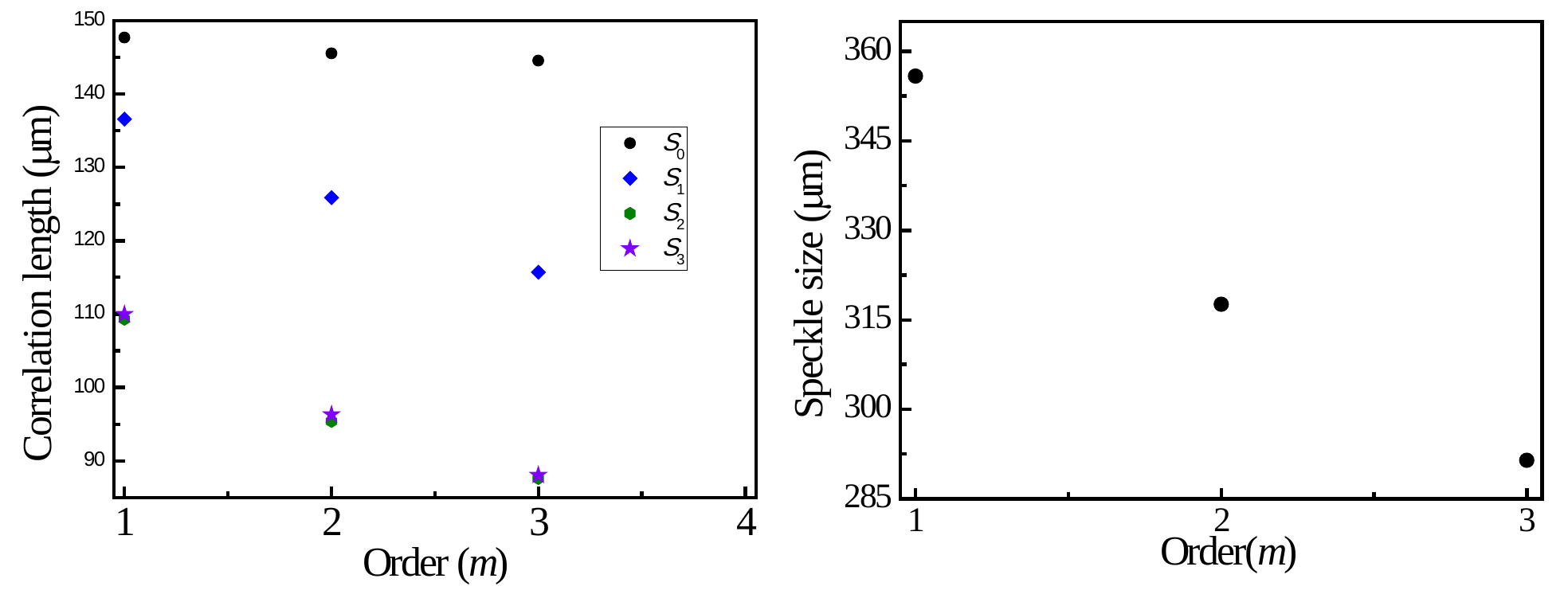}
\caption{ (Colour online) The correlation lengths for Stokes parameters (left) and the size of polarization speckles (right) vs order $m$.}\label{fig:fig8}
\end{center}
\end{figure}

Figure \ref{fig:fig8} shows the correlation lengths corresponding to the Stokes parameters corresponding to different indices  $m$=1-3. It is clear from the figure that the correlation length for $S_0$ is nearly independent of index and the correlation lengths for the remaining three Stokes parameters decrease with the index. It is very interesting to see that the scalar speckle size is independent of index. This is in contrast to the behaviour of speckles generated by the scalar vortex beams where the speckle size decreases with the increase in order as discussed in ref.\cite{vector20}. For the present case, the size of polarization speckle behaves similarly, decreasing with the increase in index. This may find applications in imaging and communications as one can control the correlation in polarization while keeping the intensity correlation the same.

\section{Conclusions}

In conclusion, we have experimentally generated the Poincar\'e beams using the polarization sensitive SLM. We have used their scattering for generating the polarization speckles i.e. spatially random polarization profile. We verified the presence of random polarization using the Stokes parameter analysis. We also showed that the size of the intensity speckles is independent of the index of the vector beam whereas the polarization speckle size decreases with the increase in index.

\newpage

\newpage

\pagebreak
\section*{Informational Fifth Page}

\end{document}